\documentclass[aps,prl,twocolumn,showpacs,superscriptaddress,groupedaddress]{revtex4}  % for review and submission
\usepackage{graphicx}  % needed for figures
\usepackage{dcolumn}   % needed for some tables
\usepackage{bm}        % for math
\usepackage{amssymb}   % for math
\usepackage{epsfig}
\usepackage{epstopdf}
\usepackage{subfigure}
\usepackage{mdframed}
\usepackage{amsmath}
\begin{document}

% The following information is for internal review, please remove them for submission

% the following line is for submission, including submission to the arXiv!!
%\hspace{5.2in} \mbox{Fermilab-Pub-04/xxx-E}

\title{Non-extremal branes}
\author{Pablo Bueno}
\author{Tom\'as Ort\'in} \affiliation{Instituto de F\'isica Te\'orica UAM/CSIC, Madrid}
\author{C. S. Shahbazi} \affiliation{Institut de Physique Th\'eorique, CEA Saclay, \^Ile-de-France}

%\input author_list.tex       % D0 authors (remove the first 3 lines
                             % of this file prior to submission, they
                             % contain a time stamp for the authorlist)
                             % (includes institutions and visitors)
\date{\today}

\begin{abstract}

We prove that for arbitrary black brane solutions of generic Supergravities there is an adapted system of variables in which the equations of motion are exactly invariant under electric-magnetic duality, \emph{i.e.} the interchange of a given extended object by its electromagnetic dual. We obtain thus a procedure  to automatically construct the electromagnetic dual of a given brane without needing to solve any further equation. We apply this procedure to construct the non-extremal $(p,q)$-string of Type-IIB String Theory (new in the literature), explicitly showing how the dual $(p,q)$-five-brane automatically arises in this construction. In addition, we prove that the system of variables used is suitable for a generic characterization of every double-extremal Supergravity brane solution, which we perform in full generality.
\end{abstract}

\maketitle
%%%%%%%%%%%%%%%%%%%%%%%%%%%%%%%%%%%%%%%%%%%%%%%%%%%%%%%%%%%%%%%%%%%%%%
%%%%%%%%%%%%%%%%%%%%%%%%%%%%%%%%%%%%%%%%%%%%%%%%%%%%%%%%%%%%%%%%%%%%%%
%%%%%%%%%%%%%%%%%%%%%%%%%%%%%%%%%%%%%%%%%%%%%%%%%%%%%%%%%%%%%%%%%%%%%%
%%%%%%%%%%%%%%%%%%%%%%%%%%%%%%%%%%%%%%%%%%%%%%%%%%%%%%%%%%%%%%%%%%%%%%
% \section{\label{sec:level1}Introduction}
% sections are not used for PRL papers
%%%%%%%%%%%%%%%%%%%%%%%%%%%%%%%%%%%%%%%%%%%%%%%%%%%%%%%%%%%%%%%%%%%%%%
%%%%%%%%%%%%%%%%%%%%%%%%%%%%%%%%%%%%%%%%%%%%%%%%%%%%%%%%%%%%%%%%%%%%%%
%%%%%%%%%%%%%%%%%%%%%%%%%%%%%%%%%%%%%%%%%%%%%%%%%%%%%%%%%%%%%%%%%%%%%%
%%%%%%%%%%%%%%%%%%%%%%%%%%%%%%%%%%%%%%%%%%%%%%%%%%%%%%%%%%%%%%%%%%%%%%
% \section{\label{sec:level1}Effective EOM for general black-branes}
Supergravity branes have played a role of outermost importance in String Theory since they were discovered to be the macroscopic counterparts of many String Theory microscopic extended objects, during the second String Revolution \cite{Polchinski:1995mt}. However, strictly speaking, this correspondence is limited to the extremal cases, which have been thoroughly studied in the literature. Much less attention has been paid to non-extremal Supergravity branes (which are regular in general, in contrast to the extremal ones), since they do not obey first order differential equations and its String Theory interpretation is less clear. In this note we are interested in further understanding the structure of general non-extremal Supergravity branes and its behaviour under \emph{electric-magnetic} duality.

In reference \cite{deAntonioMartin:2012bi}, a generalization of the FGK-formalism \cite{Ferrara:1997tw} to an arbitrary number of space-time dimensions $d$ and worldvolume dimensions $($p$+1)$ was presented. The $d$-dimensional class of theories considered in \cite{deAntonioMartin:2012bi} describes gravity coupled to a given number of scalars $\phi^{i}\, , i = 1,\dots,n_{\phi},$ and $(\text{p}+1)$-forms $A^{\Lambda}_{(\text{p}+1)}\, , \Lambda = 1,\dots,n_{A},$ and are given by the following, two-derivative, action
\begin{eqnarray}
\label{eq:daction}
S
&=&
\int d^{d}x \sqrt{|\mathrm{g}|}
\left\{
R + \mathcal{G}_{ij} (\phi)\partial_{\mu} \phi^{i} \partial^{\mu} \phi^{j}\right. \\ \notag 
&+&\left.4 \tfrac{(-1)^{\text{p}}}{(\text{p}+2)!} I_{\Lambda \Omega}(\phi) F_{(\text{p}+2)}^{\Lambda} \cdot F_{(\text{p}+2)}^{\Omega}
\right\}\, ,
\end{eqnarray}
\noindent
where $F^{\Lambda}_{(\text{p}+2)} = (\text{p}+2)d A^{\Lambda}_{(\text{p}+1)}$ are the $(\text{p}+2)$-form field strengths and the scalar dependent, negative definite, matrix $I_{\Lambda\Omega}\left(\phi\right)$ describes the couplings of scalars $\phi^{i}$ to the $(\text{p}+1)$-forms $A^{\Lambda}_{(\text{p}+1)}$. The generic space-time metric considered in \cite{deAntonioMartin:2012bi} was
\begin{eqnarray}
\label{eq:generalmetric1}
ds_{(d)}^{2}
&=&
e^{\frac{2}{\text{p}+1}U}
\left[
W^{\frac{\text{p}}{\text{p}+1}} dt^{2}
-W^{-\frac{1}{\text{p}+1}}d\vec{z}^{\, 2}_{(\text{p})}
\right]\\ \notag 
&-&e^{-\frac{2}{\tilde{\text{p}}+1}U}
\gamma_{(\tilde{\text{p}}+3)}\, ,
\end{eqnarray}
\begin{equation}
\label{eq:backgroundtransversemetric}
\gamma_{(\tilde{\text{p}}+3)}
=
\mathcal{X}^{\frac{2}{\tilde{\text{p}}+1}}
\left[
\mathcal{X}^2
\frac{d\rho^2}{(\tilde{\text{p}}+1)^2}
+ d\Omega^{2}_{(\tilde{\text{p}}+2)}
\right]\, ,	
\end{equation}

\noindent
where $\mathcal{X}\equiv \left(\frac{ \omega/2}{\sinh{\left(\frac{\omega}{2} \rho\right)}} \right)$, $\vec{z}_{(\text{p})} \equiv \left( z^{1},\dots,z^{\text{p}}\right)$ are spatial worldvolume coordinates and $d=\text{p}+\tilde{\text{p}}+4$ so $\tilde{\text{p}}$ is the number of spatial dimensions of the dual brane. $d\Omega^{2}_{(\tilde{\text{p}}+2)}$ stands for the round metric on the $(\tilde{\text{p}}+2)$-sphere of unit radius, and $\omega$ is a constant that corresponds to the \emph{non-extremality} parameter of the black-brane solution. In other words, the black-brane is extremal if and only if $\omega = 0$. 

Assuming the space-time background (\ref{eq:generalmetric1}) and that all the fields of the theory depend exclusively on the radial coordinate $\rho$, the equations of motion of (\ref{eq:daction}) are equivalent to the following set of ordinary differential equations \cite{deAntonioMartin:2012bi}
\begin{eqnarray}
\label{eq:1}
\ddot{U} +e^{2U}V_{\rm BB}
& = &
0\, ,
\\
& & \nonumber \\
\label{eq:2}
\ddot{\phi}^{i} +\Gamma_{jk}{}^{i}\dot{\phi}^{j}\dot{\phi}^{k}
+\tfrac{d-2}{2(\tilde{\text{p}}+1)(\text{p}+1)}e^{2U}\partial^{i} V_{\rm BB}
& = &
0\, ,
\\
& & \nonumber \\
\label{eq:hamiltonianconstraint}
(\dot{U})^{2}
+\tfrac{(\text{p}+1)(\tilde{\text{p}}+1)}{d-2} \mathcal{G}_{ij} \dot{\phi}^{i} \dot{\phi}^{j}
+e^{2 U} V_{\rm BB}
&  = &
c^{2}\, ,
\end{eqnarray}

\noindent
where $V_{\rm BB}$ stands for the so-called \emph{black-brane} potential
\begin{equation}\label{VBB}
V_{\rm BB}\left(\phi, q\right)\equiv 2\alpha^2\frac{2(\text{p}+1)(\tilde{\text{p}}+1)}{(d-2)} \left(I^{-1}\right)^{\Lambda\Omega} q_{\Lambda} q_{\Omega}\, ,
\end{equation}

\noindent
and $c^2$ is a real semi-definite positive constant given by
\begin{equation}\label{c}
c^2 \equiv \frac{(\text{p}+1)(\tilde{\text{p}}+2)}{4(d-2)}\omega^2 - \frac{(\tilde{\text{p}}+1)\text{p}}{4(d-2)}\gamma^2\, ,
\end{equation}

\noindent
and $\gamma$ is another constant whose origin will be clear in a moment. Notice that the system of differential equations above only involves the metric factor $U$ and the scalar fields $\phi^{i}$, since the $(\text{p}+1)$-forms can be eliminated in terms of the corresponding charges $q_{\Lambda}\, , \Lambda = 1,\dots,n_{A},$ by explicitly integrating the Maxwell equations. 
%The result is

%\begin{equation}
%\Psi^{\Lambda }
%=
%\alpha\,\int d\tau e^{2U}\left(I^{-1}\right)^{\Lambda\Omega} q_{\Omega}\, , \, \, \, \,\text{where}\, \, \, \, A^{\Lambda}_{(\text{p}+1)}
 %\equiv 
%\Psi^{\Lambda}(\rho)\, dt\wedge dz^{1} \wedge \cdots \wedge dz^{\text{p}}\, .
%\end{equation}

\noindent
Remarkably enough, it turns out that $W$ can also be explicitly integrated yielding
\begin{equation}
W = e^{\gamma\rho}\, ,
\end{equation}
\noindent
where $\gamma$ is the (integration) constant which appears in (\ref{c}).

In \cite{deAntonioMartin:2012bi} it was argued that in order to have a regular black-brane solution, we must have \footnote{In the ansatz at hand, the event horizon (if any) will correspond to $\rho\rightarrow +\infty$, whereas spatial infinity will be at $\rho  \rightarrow 0^+$. In order for the worldvolume metric to be regular in the near horizon limit, $e^U\propto e^{\frac{\omega \rho}{2}}$ and $W\sim e^{\omega \rho}$, which fixes $\gamma = \omega$.} $\gamma = \omega$ and therefore $c^2 = \frac{\omega^2}{4}$.

To sum up, in reference \cite{deAntonioMartin:2012bi} it was found that the above ansatz
%\begin{equation}
%\begin{array}{rcl}
%ds_{(d)}^{2}
%& = &
%e^{\frac{2}{\text{p}+1}U}
%\left[
%e^{\frac{\text{p}}{\text{p}+1}\omega\rho} dt^{2}
%-e^{-\frac{1}{\text{p}+1}\omega\rho}d\vec{z}^{\, 2}_{(\text{p})}
%\right]
%-
%e^{-\frac{2}{\tilde{\text{p}}+1}U}
%\gamma_{(\tilde{\text{p}}+3)\, \underline{m}\underline{n}}  dx^{m} dx^{n}\, ,
%\vspace{0.2cm}\\ \vspace{0.1cm}
%A^{\Lambda}_{(\text{p}+1)}
%& = &
%\Psi^{\Lambda}(\rho)\, dt\wedge dz^{1} \wedge \cdots \wedge dz^{\text{p}}\, ,
%\hspace{1cm}
%\dot{\Psi}^{\Lambda }
%=
%\alpha e^{2U}\left(I^{-1}\right)^{\Lambda\Omega} q_{\Omega}\, ,
% \\
%\phi^{i}
%& = &
%\phi^{i}(\rho)\, ,
%\end{array}
%\end{equation}
%\noindent 
corresponds to a black-brane solution (not necessarily regular) of the theories defined by the generic action (\ref{eq:daction}) if equations (\ref{eq:1}), (\ref{eq:2}) and (\ref{eq:hamiltonianconstraint}) are satisfied.
It can be seen that the FGK system of equations is completely fixed once we know the following data: the Riemannian metric $\mathcal{G}_{ij}$ of the non-linear sigma model, the number p of spatial dimensions of the brane and the matrix $I_{\Lambda\Omega}$ describing the couplings of the scalars and the (p+1)-forms. Actually, the FGK-system is invariant under the interchange
\begin{equation}
\label{eq:interchange}
\mathrm{p}\leftrightarrow \tilde{\mathrm{p}}\, ,
\end{equation}
\noindent
which however does not leave invariant the space-time metric, which represents now the metric of a $\tilde{\mathrm{p}}$ brane. A $\tilde{\mathrm{p}}$ brane naturally couples to a $(\tilde{\mathrm{p}}+1)$-form, that is, to the magnetic duals of the electric (p+1)-forms $A^{\Lambda}_{(\mathrm{p}+1)}$. Therefore, in order to properly perform the interchange \eqref{eq:interchange} we also have to change the electric matrix $I_{el}$ of couplings to the magnetic $I_{mag}$ one. Schematically the transformation is
\begin{equation}
\label{eq:interchangeII}
\mathrm{p}\leftrightarrow \tilde{\mathrm{p}}\, , \qquad I_{el}\leftrightarrow I_{mag}\, .
\end{equation}

\noindent
The only term in the FGK-system that depends on $I_{\Lambda\Omega}$ is the black-brane potential $V_{BB}$. Therefore, if 
\begin{equation}
\label{eq:Vbbcondition}
\left(I^{-1}\right)^{\Lambda\Omega}_{el} q_{\Lambda} q_{\Omega} = \left(I^{-1}\right)^{\Lambda\Omega}_{mag} q^{\prime}_{\Lambda} q^{\prime}_{\Omega}\, ,
\end{equation}

\noindent
where $q^{\prime}_{\Lambda} = A_{\Lambda}^{\Omega}q_{\Omega},\,\, A\in $Gl$(n_{A},\mathbb{R})$, then the FGK-system is invariant under the transformation \eqref{eq:interchangeII}, up to a redefinition of the charges, and therefore with the same solution of the FGK-system we can construct two space-time solutions, the electric-brane solution and the magnetic-brane solution. In order to see when condition \eqref{eq:Vbbcondition} holds, we have to change from electric variables $A^{\Lambda}_{(\mathrm{p}+1)}$ to the magnetic ones $\tilde{A}_{(\tilde{\mathrm{p}}+1)\Lambda}$ in the action \eqref{eq:daction}. The equations of motion and the Bianchi identities for the electric fields $A^{\Lambda}_{(\mathrm{p}+1)}$ are 
\begin{equation}
d\left(I_{\Lambda\Omega}\ast F^{\Omega}_{(\mathrm{p}+2)}\right) = 0\, , \qquad dF^{\Lambda}_{(\mathrm{p}+2)} = 0\, .
\end{equation}

\noindent
Now we define 
\begin{equation}
\label{eq:GF}
G_{(\tilde{\mathrm{p}}+2)\Lambda} = I_{\Lambda\Omega}\ast F^{\Omega}_{(\mathrm{p}+2)}\, .
\end{equation}

\noindent
and thus the equations of motion for the electric vector fields can be written as a Bianchi identity for $G_{(\tilde{\mathrm{p}}+2)\Lambda}$
\begin{equation}
dG_{(\tilde{\mathrm{p}}+2)\Lambda} = 0\Rightarrow G_{(\tilde{\mathrm{p}}+2)\Lambda} = d\tilde{A}_{(\tilde{\mathrm{p}}+1)\Lambda}\,\,\, \mathrm{locally} \, .
\end{equation}

\noindent
Equation \eqref{eq:GF} can be inverted as follows
\begin{equation}
\label{eq:GFinverted}
F^{\Lambda}_{(\mathrm{p}+2)} = (-1)^{(d-1)+(\mathrm{p}+2)(\tilde{\mathrm{p}}+2)}\left( I^{-1}\right)^{\Lambda\Omega} \ast G_{(\tilde{\mathrm{p}}+2)\Omega}
\end{equation}

\noindent
Substituting equation \eqref{eq:GFinverted} in equation \eqref{eq:daction}, we deduce that
\begin{equation}
\label{eq:LIelmag}
I_{mag} = I^{-1}_{el}\, .
\end{equation}

\noindent
Given equation \eqref{eq:LIelmag} and equation \eqref{eq:Vbbcondition} we obtain that a sufficient condition to obtain the same FGK-system for electric and magnetic branes is that there exists a matrix $A\in$ Gl$(n_{A},\mathbb{R})$ such that the following \emph{self-duality} condition holds
\begin{equation}
\label{eq:selduality}
I^{-1} = A I A^{T}\, .
\end{equation}

\noindent
Without invoking supersymmetry we can say little more beyond equation \eqref{eq:selduality}, since the couplings in the action \eqref{eq:daction} are in principle arbitrary aside from some regularity conditions. Supersymmetry, however, constrains the couplings and therefore it is easier to analyze when equation \eqref{eq:selduality} is satisfied.

Supergravity non-linear sigma models are constrainted by supersymmetry and related to the couplings of the (p+1)-forms and the scalars of the theory. Let us now consider the general situation of an extended ungauged Supergravity, where the scalar manifold is a homogeneous space of the form
\begin{equation}
\mathcal{M}_{S} = \frac{G}{H}\, ,
\end{equation}

\noindent
and the matrix $I$ of the couplings between the (p+1)-forms and the scalars is a coset representative, namely $I\in \frac{G}{H}$. The coset element $I$ must be taken in a particular representation, namely $I$ is in the representation $R(G)$ that acts on the charges of the corresponding electric p-forms of the theory. This is the standard situation happening in an extended Supergravity in diverse dimensions. From the self-duality condition \eqref{eq:selduality} we are interested in coset representatives $I$ such that there exists a matrix $A\in Gl(n_{A},\mathbb{R})$ satisfying
\begin{equation}
\label{eq:seldualityII}
I^{-1} = A I A^{T}\, .
\end{equation}

\noindent
There is a sufficient condition on $G$ such that the self-duality condition \eqref{eq:seldualityII} is implied. Let us assume that the Lie group leaves invariant a bilinear form $\mathcal{B}\in V^{\ast}\otimes V^{\ast}$, where $V$ is the $n_{A}$-dimensional representation vector space of $G$, or in other words, $q_{\Lambda}\in V$. The condition of $G$ leaving invariant $\mathcal{B}$ can be rewritten as follows
\begin{equation}
\label{eq:RB}
R^{T}\mathcal{B}R = \mathcal{B}\, ,\qquad R\in R(G)\, ,
\end{equation}

\noindent
where $R(G)$ is the corresponding representation of $G$ as automorphisms of $V$. Now, the self-duality condition does not have to be satisfied by an arbitrary element in $G$ but for an element in $G/H$ which, in the representation $R(G)$ must be symmetric in order to be an admissible $I$ \footnote{Even if it is non-symmetric, when contracting with $F^{\Lambda}_{(\mathrm{p}+2)}$ in \eqref{eq:daction} only the symmetric part survives.}. Assuming then that $R^{T} = R$ we can rewrite \eqref{eq:RB} as follows
\begin{equation}
\label{eq:RBII}
R^{-1} = \mathcal{B}^{-1} R\mathcal{B} \, ,\qquad R\in R(G)\, ,
\end{equation}
\noindent
and therefore if
\begin{equation}
\label{eq:BTB1}
\mathcal{B}^{T} = \mathcal{B}^{-1}\, ,
\end{equation}
\noindent
then equation \eqref{eq:seldualityII} is satisfied and the corresponding FGK model is self-dual, meaning that the system of differential equations to be solved for the electric p-brane and the corresponding magnetic $\tilde{\mathrm{p}}$-brane is exactly the same. 

There are several Supergravities where condition \eqref{eq:BTB1} holds. Just to name a few: Type-IIB Supergravity, where $G=$Sl$(2,\mathbb{R})$, $H = $SO$(2)$ so $\mathcal{B} = \text{antidiag}(1,-1)$; nine-dimensional $\mathcal{N}=2$ Supergravity, where $G = $Sl$(2,\mathbb{R})\times $O$(1,1)$ and $H= $O$(2)$, quotienting only the first factor and $\mathcal{B} = \text{antidiag}(1,-1)\times\text{diag}(1,-1)$; four-dimensional $\mathcal{N}=8$ Supergravity, where $G=$E$_{7(7)}$ acting on the {\bf 56} irrep. on the charges, $H = $SU$(8)/\mathbb{Z}_2$ and $\mathcal{B} $ is the symplectic form in the ${\bf 56}$-dimensional vector space; four-dimensional $\mathcal{N}=6$ Supergravity, with  $G=$SO$^*(12)$, $H = $U$(6)$ and $\mathcal{B}$ is the identity matrix, etc.

%%%%%%%%%%%%%%%%%%%%%%%%%%%%%%%%%%%%%%%%%%%%%%%%%%%%%%%%%%%%%%%%%%%%%%
%%%%%%%%%%%%%%%%%%%%%%%%%%%%%%%%%%%%%%%%%%%%%%%%%%%%%%%%%%%%%%%%%%%%%%
%%%%%%%%%%%%%%%%%%%%%%%%%%%%%%%%%%%%%%%%%%%%%%%%%%%%%%%%%%%%%%%%%%%%%%
%%%%%%%%%%%%%%%%%%%%%%%%%%%%%%%%%%%%%%%%%%%%%%%%%%%%%%%%%%%%%%%%%%%%%%
%\section{The $(p,q)$-black-string of Type-IIB Supergravity}
%\label{sec:FGKstring}
%%%%%%%%%%%%%%%%%%%%%%%%%%%%%%%%%%%%%%%%%%%%%%%%%%%%%%%%%%%%%%%%%%%%%%
%%%%%%%%%%%%%%%%%%%%%%%%%%%%%%%%%%%%%%%%%%%%%%%%%%%%%%%%%%%%%%%%%%%%%%
%%%%%%%%%%%%%%%%%%%%%%%%%%%%%%%%%%%%%%%%%%%%%%%%%%%%%%%%%%%%%%%%%%%%%%
%%%%%%%%%%%%%%%%%%%%%%%%%%%%%%%%%%%%%%%%%%%%%%%%%%%%%%%%%%%%%%%%%%%%%%
Let us see how this works in an particular example, namely the $(p,q)$-black-strings and $(p,q)$-5-black branes of Type-IIB Supergravity. First, we will use the effective FGK variables to construct the non-extremal $(p,q)$-black-string, new in the literature, and then, we will show how in the FGK framework this solution is actually the same as the non-extremal $(p,q)$-5-black-brane, also new. Before getting started, let us review the basic properties of the extremal $(p,q)$-string of Schwarz \cite{Schwarz:1995dk}.

From the stringy perspective, a (extremal) $(p,q)$-string is a bound state of Type-IIB String Theory composed of $p$ \emph{D-strings} (\emph{D1s}), charged under the RR two-form $C_{(2)}$, and $q$ \emph{fundamental strings} (\emph{F1s}), with charge under the NS-NS two-form $B$. Type-IIB Supergravity is invariant under a global SL$(2,\mathbb{R})$ symmetry, so all the states of the theory are accomodated in multiplets of such group. In particular any state can be generated from another one living in the same multiplet by applying a SL$(2,\mathbb{R})$ transformation. This is the case for the $D1$ and $F1$ solutions, which are related to each other via this IIB S-duality. Similarly, we can generate a $(p,q)$-string starting from one of them, and performing a general enough SL$(2,\mathbb{R})$ transformation. This was done for the first time by Schwarz \cite{Schwarz:1995dk}, who also gave the corresponding Supergravity version of the solution. In fact, from the Supergravity perspective, all these states correspond to extremal black strings charged under one or both two-forms. All these solutions are nevertheless singular, given that the corresponding black-string singularities are naked. As we will see, this behavior is cured in the non-extremal case, and we will be able to construct a regular non-extremal $(p,q)$-black-string solution. 
The relevant truncated Type-IIB Supergravity Lagrangian is
\begin{eqnarray}
  \label{eq:IIBactionEtruncated}
  S  =  \int \, d^{10} x \sqrt{|\mathrm{g}|}\,\left[R + \frac{1}{2}\frac{\partial_{\mu}\tau\partial^{\mu}\bar{\tau}}{\left(\Im{\rm m}\tau\right)^2} + \frac{1}{2\cdot 3!}\mathcal{H}^{T}\mathcal{M}^{-1} \mathcal{H} \right]\,  \,\,
\end{eqnarray}

\noindent
where $\mathcal{H}\equiv d\mathfrak{B}$, with $\mathfrak{B}^T\equiv \left(C_{(2)},B \right)$ and $\mathcal{M}\equiv \frac{1}{\Im \mathfrak{m} \tau}\left( \begin{array}{cc}
|\tau|^2 & \Re \mathfrak{e} \tau  \\
\Re \mathfrak{e} \tau & 1  \end{array} \right)$ with $\Im \mathfrak{m} \tau>0$ is the coset representative of the space SL$(2,\mathbb{R})/$SO$(2)$ parametrized by the axidilaton $\tau\equiv C_{(0)}+ie^{-\Phi}$. Since black strings in ten dimensions have $\text{p}=1$ and $\tilde{\text{p}}=5$, let us set \footnote{There should be no confusion about the p that denotes the number of spatial dimensions of a given brane and the $p$ in the $(p,q)$-strings, which corresponds to its charge under $C_{(2)}$.}
\begin{equation}\label{pene}
d=10\, , \quad \text{p}=1\, , \quad \tilde{\text{p}} = 5\, 
\end{equation}
in the FGK effective action (\ref{eq:daction}).
%\begin{equation}
%\label{eq:10action}
%S[g,A^{\Lambda}_{(2)},\phi^{i}]
%=
%\int d^{10}x \sqrt{|\mathrm{g}|}
%\left\{
%R + \mathcal{G}_{ij} (\phi)\partial_{\mu} \phi^{i} \partial^{\mu} \phi^{j}
%- \tfrac{2}{3} I_{\Lambda \Omega}(\phi) F_{(3)}^{\Lambda} \cdot F_{(3)}^{\Omega}
%\right\}\, .
%\end{equation}
%\noindent
%In addition, (\ref{eq:generalmetric1}) becomes in this case
%\begin{equation}
%\label{eq:generalmetric10}
%ds_{(10)}^{2}
%=
%e^{U}
%\left[
%e^{\frac{1}{2}\omega\rho} dt^{2}
%-e^{-\frac{1}{2}\omega\rho}dz^{2}
%\right]
%-
%e^{-\frac{1}{3}U}
%\gamma_{(8)\, \underline{m}\underline{n}}  dx^{\underline{m}} dx^{\underline{n}}\, ,
%\end{equation}
%\begin{equation}
%\label{eq:backgroundtransversemetric10}
%\gamma_{(8)\, \underline{m}\underline{n}}  dx^{m} dx^{n}
%=
%\left(\frac{ \omega/2}{\sinh{\left(\frac{\omega}{2} \rho\right)}} \right)^{\frac{1}{3}}
%\left[
%\left( \frac{\omega/2}{\sinh{\left(\frac{\omega}{2} \rho \right)}}\right)^2
%\frac{d\rho^2}{6^2}
%+ d\Omega^{2}_{(7)}
%\right]\, ,	
%\end{equation}
%\noindent
%and the equations of motion are given by
%\begin{eqnarray}
%\label{eq:110}
%\ddot{U} +e^{2U}V_{\rm BB}
%& = &
%0\, ,
%\\
%& & \nonumber \\
%\label{eq:210}
%\ddot{\phi}^{i} +\Gamma_{jk}{}^{i}\dot{\phi}^{j}\dot{\phi}^{k}
%+\frac{1}{3}e^{2U}\partial^{i} V_{\rm BB}
%& = &
%0\, ,
%\\
%& & \nonumber \\
%\label{eq:hamiltonianconstraint10}
%(\dot{U})^{2}
%+\tfrac{3}{2} \mathcal{G}_{ij} \dot{\phi}^{i} \dot{\phi}^{j}
%+e^{2 U} V_{\rm BB}
%&  = &
%c^{2}\, .
%\end{eqnarray}
\noindent
Now, the key point to notice is that the action (\ref{eq:IIBactionEtruncated}) is a particular case of (\ref{eq:daction}), by taking $n_{\phi}=2\, , n_{A} = 2$ and making the following identifications
\begin{equation}
\label{eq:identificationfields}
\phi^{1} = C_{(0)}\, , \, \phi^{2} =e^{-\Phi}\, , \, \mathcal{G}_{ij} =e^{2\Phi} \frac{\delta_{ij}}{2}\, , \, I(\phi) \equiv -\frac{1}{8}\mathcal{M}^{-1}\, ,
\end{equation}

\noindent
where $i,j =1,2$ and $\tau = C_{(0)}+i e^{-\Phi}$. We thus obtain that the black-brane potential for this truncation of Type-IIB Supergravity is given by
\begin{equation}
\label{eq:blackstringsV}
-V_{\text{BB}}\left(\phi, q\right) = \mathcal{M}^{\Lambda\Omega} q_{\Lambda} q_{\Omega} = e^{\Phi}\left( \left|\tau\right|^2 p^2 +q^{2} + 2 pqC_{(0)}\right)\, ,
\end{equation}
\noindent
where $\Lambda, \Omega = 1,2$ and we have defined $\alpha^2 = \frac{1}{2^4\cdot 3}$ and $q_1\equiv p$, $q_2\equiv q$. Therefore, in order to obtain the black-string solutions of the theory (\ref{eq:IIBactionEtruncated}) we \emph{just} have to solve the system of ordinary differential equations given by (\ref{eq:1}), (\ref{eq:2}) and (\ref{eq:hamiltonianconstraint}) assuming equations (\ref{pene}), (\ref{eq:identificationfields}) and (\ref{eq:blackstringsV}). Notice that $\mathcal{M}$ is definite positive and therefore $V_{\text{BB}}\left(\phi, q\right)$ in (\ref{eq:blackstringsV}) is negative definite.

In reference \cite{deAntonioMartin:2012bi}, it was shown that for regular extremal black-brane solutions, the value $\phi_{H}$ of the scalars at the black-brane horizon obeys
\begin{equation}
\label{eq:BBattractorsIiI}
\partial_{i} V_{\text{BB}}\left(\phi_{H}, q\right) = 0\, ,\qquad i = 1,\dots, n_{\phi}\, .
\end{equation}
The solutions $\phi_{H}$ of equation (\ref{eq:BBattractorsIiI}) are the so-called black-brane attractors, and generalize to black-brane solutions the popular concept of black-hole attractor. Notice that equation (\ref{eq:BBattractorsIiI}) completely fixes the value of the scalars at the horizon in terms of the charges, as long as there are no \emph{flat directions}. Taking the black-brane potential as in (\ref{eq:blackstringsV}), one easily finds that (\ref{eq:BBattractorsIiI}) has no solutions for the $(p,q)$-black-string system, meaning that there does not exist any extremal regular black-string solution of Type-IIB Supergravity with non-trivial scalars.

The most general extremal solution of this kind was constructed by Schwarz in \cite{Schwarz:1995dk}. It is given, in standard coordinates by
\begin{eqnarray}
\label{eq:pqs}
&ds_{E}^2 = H^{-\frac{3}{4}} \left[ dt^2 - dz^2\right] - H^{\frac{1}{4}}d\vec{x}^2\, , \\ \notag
&\mathfrak{B}_{tz}=\mathfrak{a}\left(H^{-1}-1 \right)\, , \, \mathcal{M} = \mathfrak{a} \mathfrak{a}^T H^{-\frac{1}{2}} + \mathfrak{b}\mathfrak{b}^T H^{\frac{1}{2}}\, ,\,
\end{eqnarray}
where 
\begin{equation}
 \displaystyle H = 1 + \frac{h}{r^6}\, , \,
\end{equation}
 $r^2 \equiv \vec{x}^2$ and $\mathfrak{a}^T = (a_{1}, a_{2})$ and $\mathfrak{b}^T = (b_{1}, b_{2})$ are two constant vectors to be expressed in terms of the physical parameters of the solution and subject to the constraint $\mathfrak{a}^T\eta \mathfrak{b}=a_1 b_2-a_2 b_1=1$. The relation between $\mathcal{M}$ and $H$ can be inverted to obtain the expression for the axidilaton, which reads
\begin{equation}
\tau = \frac{a_{1} a_{2}  + b_{1} b_{2} H}{a^2_{2}  + b^2_{2} H} + \frac{i \sqrt{H}}{a^2_{2} + b^2_{2} H}\, .
\end{equation}
It is not difficult to recover the $D1$ and $F1$ solutions from the $(p,q)$-black-string one by setting $C_{(0)}=0$ and $q=0$ or $p=0$ respectively in each case.

The standard coordinates can be related to the FGK ones through the change $r=\rho^{-\frac{1}{6}}$. It is straightforward to check that equations (\ref{eq:1}), (\ref{eq:2}) and (\ref{eq:hamiltonianconstraint}) with $c=0$ are satisfied by Schwarz's $(p,q)$-black-string (\ref{eq:pqs}) \footnote{In particular, the relation between $U(\rho)$ and $H(r)$ is given by $H(r)^{-\frac{3}{4}}=e^{U(\rho)}$.}.
%We can start recovering the well-known extremal solutions. In order to do so, notice that the extremal limit of (\ref{eq:generalmetric10}) is 
%
%\begin{equation}
%\label{eq:generalmetric15}
%ds_{(10)}^{2}
%=
%e^{U}
%\left[
%dt^{2}
%-dz^{2}
%\right]
%-
%\left(e^{U}\rho\right)^{-\frac{1}{3}} 
%\left[
%\frac{d\rho^2}{6^2\rho^2}
%+ d\Omega^{2}_{(7)}
%\right]\, ,
%\end{equation}
%so the usual black-brane coordinates appearing in (\ref{eq:susystringmetric1}) can be related to the FGK ones through the change of variables
%\begin{equation}\label{coor}
%r=\rho^{-\frac{1}{6}}\, ,
%\end{equation}
%and the identification
%\begin{equation}\label{coor2}
%H(r)^{-\frac{3}{4}}=e^{U(\rho)}\, .
%\end{equation}
%
%\noindent
%It is straightforward to check that equations (\ref{eq:110}), (\ref{eq:210}) and (\ref{eq:hamiltonianconstraint10}) with $c=0$ are satisfied by Schwarz's $(p,q)$-black-string (\ref{pqext}). 
We find that the singular extremal $(p,q)$-black-string can be generalized to a regular non-extremal solution, given by
\begin{widetext}
\begin{eqnarray}
\label{eq:susystringmetric14}
ds_{E}^2& = &H^{-\frac{3}{4}} \left[W dt^2 - dz^2\right] - H^{\frac{1}{4}} \left[W^{-1}dr^2+r^2 d\Omega^2_{(7)} \right]\, ,\\ \notag
\mathfrak{B}_{tz}&=&\pm \mathfrak{a}\left( H^{-1}-1\right)\, , \, \,\,\,   \tau=\displaystyle \frac{a_{1} a_{2}  + b_{1} b_{2} H}{a^2_{2}  + b^2_{2} H} + \frac{i \sqrt{H}}{a^2_{2} + b^2_{2} H}\, ,\\ \notag
\displaystyle H& =& 1 + \frac{h}{r^6}\, ,\,\, W=1+\frac{2c}{r^6}\, , \,\, h=c + \frac{2}{\sqrt{3}}\sqrt{|{V_{\rm BB}}_{\infty}|+\frac{3c^2}{4}}\, , 
\end{eqnarray}
\begin{eqnarray} \notag
a_1&=& \frac{\left(q\,C_{(0)\infty}+p|\tau_{\infty}|^2\right)e^{\Phi_{\infty}}}{\sqrt{|{V_{\rm BB}}_{\infty}|}}\, ,\, \,b_1=-\frac{q\, }{\sqrt{|{V_{\rm BB}}_{\infty}|}}\, ,\\ \notag
a_2&=& \frac{\left(q+p\,C_{(0)\infty}\right)e^{\Phi_{\infty}}}{\sqrt{|{V_{\rm BB}}_{\infty}|}}\, ,\, \, \, b_2=\frac{p\,}{\sqrt{|{V_{\rm BB}}_{\infty}|}}\, , \\ \notag
{V_{\rm BB}}_{\infty}&\equiv& -e^{\Phi_{\infty} }\left(q^2+2pqC_{(0)\infty}+p^2|\tau_{\infty}|^2\right)\, ,
\end{eqnarray}
%\end{mdframed}
%\vspace{1cm}
\end{widetext}
\noindent
where we have expressed all the parameters of the solution in terms of the corresponding physical quantities (charges $\mathfrak{q}$ and asymptotic values of the axion and dilaton). The FGK variables in which this solution was obtained are related to the standard ones by the change of variables
\begin{eqnarray}
r^6=\frac{2c}{e^{2c\rho}-1}\, ,\,\,\, H(r)^{-3/4}=e^{U(\rho)}e^{-c\rho}\, .
\end{eqnarray}

It can be easily seen that the general non-extremal solution we have found reduces to all the known solutions, namely, the non-extremal $D1$-brane by taking $C_{(0)} = 0,\, q = 0$; the non-extremal $F1$-string by setting $C_{(0)} = 0,\, p = 0$; and Schwarz's extremal $(p,q)$-string by taking the $c\rightarrow 0$ limit.
This non-extremal $(p,q)$-black-string posseses the same metric as the non-extremal $D1$ and $F1$, and an axidilaton with both real an imaginary parts having the same expression as Schwarz's extremal $(p,q)$-string (\ref{eq:pqs}) (although everything depends now also on the non-extremality parameter $c=\omega/2$).

%\subsection{Electromagnetic duality in the FGK and non-extremal $(p,q)$-5-brane}

%%%%%%%%%%%%%%%%%%%%%%%%%%%%%%%%%%%%%%%%%%%%%%%%%%%%%%%%%%%%%%%%%%%%%%
%%%%%%%%%%%%%%%%%%%%%%%%%%%%%%%%%%%%%%%%%%%%%%%%%%%%%%%%%%%%%%%%%%%%%%

As we explained before, the FGK equations (\ref{eq:1}), (\ref{eq:2}) and (\ref{eq:hamiltonianconstraint}) are blind under electric-magnetic duality for a broad class of bosonic actions. That is indeed the case of the action (\ref{eq:IIBactionEtruncated}). Indeed, all the equations of motion of the FGK-formalism coming from (\ref{eq:IIBactionEtruncated}) are invariant under the interchange $\mathrm{p}\leftrightarrow \tilde{\mathrm{p}}\, , \,\, I_{el}\leftrightarrow I_{mag}$.
The only subtlety appears in the black-brane potential. Since $\mathcal{M}^{-1} = \eta^{T}\,\mathcal{M}\,\eta\,$, this goes from
\begin{equation}
\label{eq:blackstringsVi}
-V_{BB}^{(C_{(2)},B)}=\mathfrak{q}^T \mathcal{M} \mathfrak{q} = e^{\Phi}\left( \left|\tau\right|^2 p^2 +q^{2} + 2 pqC_{(0)}\right)\, ,
\end{equation}
\noindent
in the \emph{electric} version of the action, to
\begin{equation}
\label{eq:blackstringsViii}
-V_{BB}^{(C_{(6)},B^{(6)})}=\mathfrak{q}_5^T \mathcal{M} \mathfrak{q}_5= e^{\Phi}\left( \left|\tau\right|^2 p_5^2 +q_5^{2} + 2 p_5q_5C_{(0)}\right)\, , 
\end{equation}
in the \emph{magnetic} one, provided that we define the charges $\mathfrak{q}_5$ as 
\begin{equation}
\mathfrak{q}_5=(p_5,q_5)^T\equiv \eta \mathfrak{q} =(q,-p)^T\, , \,\,\, \eta = \left( \begin{array}{cc}
0 & 1  \\
-1 & 0  \end{array} \right)\, .
\end{equation}

%\noindent
%since
%
%\begin{equation}
%\mathcal{M}^{-1} = \eta^{T}\,\mathcal{M}\,\eta\, .
%\end{equation}
%
%\noindent
%Altogether, $V_{\rm BB}$ goes from 
%
%\begin{equation}
%\label{eq:blackstringsVi}
%-V_{BB}^{(C_{(2)},B)}\left(\tau, p,q\right) =\mathfrak{q}^T \mathcal{M} \mathfrak{q} = e^{\Phi}\left( \left|\tau\right|^2 p^2 +q^{2} + 2 pqC_{(0)}\right)\, ,
%\end{equation}

%\noindent
%in the \emph{electric} version of the action, to

%\begin{equation}
%\label{eq:blackstringsVii}
%-V_{BB}^{(C_{(6)},B^{(6)})}\left(\tau,P, Q\right) =\mathfrak{Q}^T \mathcal{M}^{-1} \mathfrak{Q} \, ,
%\end{equation}

%\noindent
%in the \emph{magnetic} version of it, where $\mathfrak{Q}\equiv(P,Q)^T$ are the charges under the $C_{(6)}$ and the $B^{(6)}$ \emph{magnetic} forms respectively. By exploting the property $\mathcal{M}^{-1}=\eta^T \mathcal{M} \eta$ it is trivial to find

%\begin{equation}
%\label{eq:blackstringsViii}
%-V_{BB}^{(C_{(6)},B^{(6)})}\left(\tau,P, Q\right) =\mathfrak{q}_5^T \mathcal{M} \mathfrak{q}_5= e^{\Phi}\left( \left|\tau\right|^2 p_5^2 +q_5^{2} + 2 p_5q_5C_{(0)}\right)\, , 
%\end{equation}

%\noindent
%where we have defined $\mathfrak{q}_5=(p_5,q_5)^T\equiv \eta \mathfrak{Q} =(Q,-P)^T$. 
Hence, in the effective FGK variables, pairs consisting of a black string and a $5$-black-brane solving the equations of motion of the corresponding ten-dimensional action appear as a single solution. This corresponds in general to a black string of charges $(p,q)$ under $(C_{(2)},B)$ and a $5$-black-brane with charges $(q,-p)$ under $(C_{(6)},B^{(6)})$. Also, the fact that both black-brane potentials are equivalent implies that no regular $5$-black-brane extremal objects exist. 

The known $5$-brane solutions of Type-IIB Supergravity correspond to the non-extremal \emph{D5-brane}, the non-extremal \emph{S5} and the analogous of Schwarz's extremal black-string, the \emph{$(p,q)$-5-brane} of Lu and Roy \cite{Lu:1998vh}. Using the very same solution of the FGK system (\ref{eq:susystringmetric14}) it is straightforward to construct the non-extremal \emph{$(p,q)$-5-brane}, which can be easily seen to reduce to the known cases just mentioned.

As we have explained, there is a black-brane attractor mechanism at work for extremal black-branes $(\omega=0)$, which fixes the scalars at the horizon as the critical points $\phi_{H}$ of the black-brane potential. Indeed, assuming regularity of the scalars at the horizon as well as a regular Riemannian scalar metric, the value of the scalars at the horizon $\phi_{H}$ for an extremal black-brane solution satisfies (\ref{eq:BBattractorsIiI}).
We will use now the FGK-formalism for black-branes to prove the existence of a universal \footnote{In the sense that it will have the same expression for any theory of the form (\ref{eq:daction}).} black-brane solution with constant scalars, and a universal near-horizon behaviour, if condition (\ref{eq:BBattractorsIiI}) is satisfied. In this case, however, such condition will appear as a constraint from imposing the scalars to be constant (often refered to as \emph{double-extremality}) and not from requiring the non-extremality parameter $c$ to vanish. Indeed, for constant scalars, the FGK system of equations reduces to
\begin{eqnarray}
\label{eq:1c}
\ddot{U} +e^{2U}V_{\rm BB}
& = &
0\, ,
 \\ 
\label{eq:2c}
\partial_{i} V_{\rm BB}
& = &
0\, ,
\\
\label{eq:hamiltonianconstraintc}
(\dot{U})^{2}+e^{2 U} V_{\rm BB} &  = &  c^{2}\, .
\end{eqnarray}

\noindent
Note that equations (\ref{eq:1c}), (\ref{eq:2c}) and (\ref{eq:hamiltonianconstraintc}) do not depend on the number p of spatial dimensions of the brane. Notice also that $V_{\rm BB}(q)$ will be now a constant constructed from the product of the constant $n_A\times n_A$ kinetic matrix $\left(I^{-1}\right)^{\Lambda\Omega}$ and the charge vectors $q_{\Lambda}$, see (\ref{VBB}). Thus, a double-extremal black brane will in general be charged under the $n_A$ $($p$+1)$-forms $A_{(\text{p}+1)}^{\Lambda}$ present in the theory. 

Equation (\ref{eq:2c}) can be automatically solved if the black-brane potential has at least one critical point, something that must be analyzed in a case by case basis and that we will assume henceforth. Equation (\ref{eq:1c}) is the derivative of equation (\ref{eq:hamiltonianconstraintc}), and thus we are left with a single equation. This was to be expected, provided there is only one variable left to be integrated, namely $U$. Equation (\ref{eq:hamiltonianconstraintc}) can be explicitly integrated and the solution is given by
\begin{equation}
\label{eq:nonextremaluniversal}
e^{-2U} =\frac{|V_{\rm BB}| \sinh^2\left(c\rho + s\right)}{c^2} \, ,
\end{equation}

\noindent
where $s$ is an integration constant. Normalizing the metric to obtain Minkowski space-time at spatial infinity fixes $s$ to be given by
\begin{equation}
\label{eq:s}
s = \mathrm{arcsinh}\left(\frac{c}{\sqrt{|V_{\rm BB}|}}\right)\, .
\end{equation}

\noindent
Therefore, inserting equation (\ref{eq:nonextremaluniversal}) into the general metric (\ref{eq:generalmetric1}) we obtain a complete $(p_1,p_2,$ $...,p_{n_A})$-p-black-brane solution with constant scalars which solves the theory (\ref{eq:daction}). The metric factor $e^{-2U}$ is well defined for $\rho\in [0,+\infty )$ and therefore the solution contains a horizon at $\rho\rightarrow +\infty$ and is regular. Taking the extremal limit $c\to 0$ we obtain
\begin{equation}
\label{eq:extremaluniversal}
e^{-2U} = \left(1+\sqrt{|V_{\rm BB}|}\rho\right)^2\, ,
\end{equation}

\noindent
which corresponds to a regular extremal universal black-brane solution. We can obtain now the near-horizon geometry of the extremal solution simply by taking the limit $\rho\to+\infty$ in the general extremal metric where now $U$ is given by equation (\ref{eq:extremaluniversal}). Making the change of coordinates $\rho = r^{\text{p}+1} $ and relabeling $\vec{z}$ and $t$ we can rewrite the final result as follows
\begin{eqnarray}
\label{eq:nearhorizonextremal2}
&&\hspace{-0.2cm}\lim_{\rho\to\infty} ds_{(d)}^{2} =\\ \notag &&\hspace{-0.2cm} |V_{\rm BB}|^{\frac{1}{\tilde{\text{p}}+1}}\left[\frac{(\text{p}+1)^2}{(\tilde{\text{p}}+1)^2}\, \frac{1}{r^2}\left[dt^{2}-d\vec{z}^{\, 2}_{(\text{p})} -  dr^2\right]
+  d\Omega^{2}_{(\tilde{\text{p}}+2)}\right]\, ,
\end{eqnarray}

\noindent
which corresponds to the space AdS$_{(2+\text{p})}\times S^{\tilde{\text{p}}+2}$. Notice that the near-horizon geometry (\ref{eq:nearhorizonextremal2}) is itself a solution of the equations of motion, and corresponds again to a universal solution with constant scalars. Let us remind the reader that in order for both the universal black-brane solution, or the near-horizon solution to exist, the only requirement is that the $n_{\phi}$ scalars present in the theory can be consistently chosen to be constant. This is equivalent to requiring the black-brane potential to have a critical point.

A simple case in which we can easily construct the double-extremal solution corresponds to $\mathcal{N}= 2$, $d=5$ supergravity coupled to one vector multiplet. A model of this theory gets completely determined by specifying a completely symmetric tensor $C_{IJK}$ (see, e.g. \cite{deAntonioMartin:2012bi}, for details), which in this case reads $C_{011}=1/3$. The black-brane potential of the model reads
\begin{equation}\label{vv}
-V_{\rm BB}=\frac{1}{3}\left[ \left(p^0\right)^2e^{-2\sqrt{\frac{2}{3}}\phi}+2\left(p^1 \right)^2e^{\sqrt{\frac{2}{3}}\phi}\right]\, ,
\end{equation}
being $\phi$ the only scalar of the theory, and $p^0$, $p^1$ the charges under the 2-forms $B_{0\mu\nu}$ and $B_{1\mu\nu}$ dual to the graviphoton and the 1-form of the vector multiplet respectively \cite{deAntonioMartin:2012bi}. Now, (\ref{vv}) has a critical point for
\begin{equation}\label{phii}
\phi_h=\sqrt{\frac{2}{3}}\log \left(\left|\frac{p^0}{p^1}\right| \right)\, ,
\end{equation}
at which
\begin{equation}\label{vvv}
-V_{\rm BB}(\phi_h,p)=\left[|p^0|(p^1)^2\right]^{2/3}\, .
\end{equation}
Therefore, the double-extremal black string of this model is given by
\begin{equation}
e^{-2U} =\frac{\left[|p^0|(p^1)^2\right]^{2/3} \sinh^2\left(c\rho + s\right)}{c^2} \, ,
\end{equation}
with 
\begin{equation}
s = \mathrm{arcsinh}\left(\frac{c}{\left[|p^0|(p^1)^2\right]^{1/3} }\right) \, .
\end{equation}

\noindent
\textbf{Acknowledgements.} This work has been supported in part by the Spanish Ministry of Science and Education grant FPA2012-35043-C02 (-01 \& -02), the Centro de Excelencia Severo Ochoa Program grant SEV-2012-0249, the Comunidad de Madrid grant HEPHACOS S2009ESP-1473 and the Spanish Consolider-Ingenio 2010 program CPAN CSD2007-00042 and EU-COST action MP1210 “The String Theory Universe”. The work was further supported by the JAE-predoc grant JAEPre 2011 00452 (PB) and the ERC Starting Independent Researcher Grant 259133, ObservableString (CSS). Research at IFT is supported by the Spanish MINECO's Centro de Excelencia Severo Ochoa Programme under grant SEV-2012-0249. TO wishes to thank M.M. Fernández for her permanent support.

\bibliographystyle{JHEP}
\bibliography{References.bib}
%\bibliography{C:/Users/cshabazi/Dropbox/Referencias/References}
\label{biblio}

\end{document}